\documentclass[aps,prl,floats,twocolumn]{revtex4}
\usepackage{graphicx}
\usepackage{color}
\begin{document}


\newcommand{\etal} {{\it et al.}}
\newcommand{\ie} {{\it i.e.}}


\title{Electronic Correlations in CoO$_2$, the Parent Compound of Triangular Cobaltates}

\author{C. de Vaulx$^1$, M.-H. Julien$^{1,*}$, C. Berthier$^{1,3}$, S. H\'ebert$^{2}$, V. Pralong$^{2}$, and A. Maignan$^{2}$}

\affiliation{$^1$ Laboratoire de Spectrom\'etrie Physique, Universit\'e J. Fourier and CNRS, BP87, 38402 Saint Martin
d'H\`{e}res, France}

\affiliation{$^2$ Laboratoire CRISMAT, UMR6508 CNRS ENSICAEN, 6 bd M.Juin, 14050 Caen Cedex 4, France}

\affiliation{$^3$ Grenoble High Magnetic Field Laboratory, CNRS, BP 166, 38042 Grenoble Cedex 9, France}

\date{\today}

\begin{abstract}

A $^{59}$Co NMR study of CoO$_2$, the $x=0$ end member of A$_x$CoO$_2$ (A=Na, Li...) cobaltates, reveals a metallic ground state,
though with clear signs of strong electron correlations: low-energy spin fluctuations develop at wave vectors $q\neq0$ and a
crossover to a Fermi-liquid regime occurs below a characteristic temperature $T^*\simeq 7$~K. Despite some uncertainty over the
exact cobalt oxidation state in this material, the results show that electronic correlations are revealed as $x$ is reduced below
0.3. The data are consistent with Na$_x$CoO$_2$ being close to the Mott transition in the $x\rightarrow 0$ limit.

\end{abstract}
\maketitle

Doping a Mott insulator (a state where electrons are forced to localize by Coulomb repulsion $U$) with charge carriers can lead
to a variety of strongly correlated electronic phases with outstanding properties such as $d$-wave superconductivity or colossal
magnetoresistance~\cite{Imada}. Recently, studies of sodium cobalt oxides Na$_x$CoO$_2$ have revealed a rich phase diagram with
superconductivity, anomalous magnetism and strong thermopower. A large number of theoretical attempts to explain these phenomena
are based on the infinite-$U$ limit of the Hubbard Hamiltonian, the $t$-$J$ model. In this approach, sodium cobaltates are
triangular analogs of (electron doped) cuprate superconductors and the "undoped" $x=0$ phase CoO$_2$ should be a Mott insulator.

The doped Mott insulator picture is, however, challenged by the experimental observation that Na$_x$CoO$_2$ appears to become a
less and less correlated metal as $x$ is reduced from $\sim$0.7 down to 0.18 (if one excepts the singular $x=0.5$
composition)~\cite{Wu06}. This raises a two-fold question: Is the ground state of CoO$_2$ metallic or insulating~\cite{theoryU}?
If metallic, how far is CoO$_2$ from the Mott transition, which is predicted to occur at $U/|t|\simeq10.5-12$ for the Hubbard
model on a triangular lattice~\cite{theory}? In other words, how strong are electron correlations in CoO$_2$?

The $x=0$ limit is not just theoretical: In 1996, Amatucci, Tarascon and Klein discovered that a well-defined (O1) phase forms
right at $x=0$ after full electrochemical delithiation of Li$_x$CoO$_2$ \cite{Amatucci,Tarascon}. This phase, which is not stable
in air, has since been reproduced by several groups~\cite{H13,P3}. From the structural point of view, the CoO$_2$ blocks are
identical to those in Li$_x$CoO$_2$ or Na$_x$CoO$_2$. The $c$-axis parameter is somewhat contracted, but the in-plane Co-Co
distance is almost unchanged. Furthermore, the stoichiometric nature of this phase is manifested in the established phase
separation with the adjacent H1-3 phase (Li$_{0.12}$CoO$_2$)~\cite{H13}. CoO$_2$ is thus the ideal candidate for studying the
strategic $x=0$ limit of triangular cobaltates.

In this Letter, we report the bulk and local magnetic properties of CoO$_2$ using magnetization and $^{59}$Co NMR data. CoO$_2$
is found to be a metal rather than an insulator. Nevertheless, the evidence for spin fluctuations and for a Fermi liquid regime
occurring at much lower temperature ($T$) than in Na$_{0.3}$CoO$_2$ highlight the vigor of electron correlation and the probable
closeness to the Mott transition.

Polycrystalline CoO$_2$ samples were obtained by oxidizing stoichiometric LiCoO$_2$ as the positive electrode material in
Swagelok type cells charged to 5.1~V. The cells were assembled in an argon filled glove box with lithium metal as the negative
electrode and a borosilicate glass fiber sheet saturated with 1~M LiPF$_6$ in sulfone (Merck) as the electrolyte. The composite
positive electrode was prepared by mixing 80~wt.\% active material with 20~wt.\% acetylene black. The X-ray diffraction (XRD)
pattern of the synthesized powder, recorded in an atmosphere-controlled chamber, displays the sharp peaks typical of crystalline
CoO$_2$ (Inset to Fig.~2a). The refined cell parameters are $a=b=2.806(1)$~\AA, $c=4.313(4)$~\AA~(space group P-3m). The peak at
$2\theta=19.65^\circ$, also reported by other groups, is attributed to the Li-rich H1-3 phase~\cite{H13}.

Samples ($\sim$40~mg) were placed under an inert atmosphere inside a sealed quartz tube, in order to prevent any exposure to air.
Degradation of CoO$_2$ was ruled out by XRD measurements performed after the NMR experiment. Reproducible NMR results were
obtained from three different samples. We have checked that the samples exposed to air showed, in contrast to CoO$_2$, a large
distribution of nuclear relaxation rates extending to much longer values, as well as spectra with significantly lower shifts and
unresolved quadrupolar structure.

$^{59}$Co NMR spectra recorded at different magnetic field values in randomly oriented powders provide a powerful way to
distinguish the two phases seen in XRD: In a field of $\sim$13.5~T, the difference between the high field side of the NMR
spectrum and the simulation with a single cobalt site (grey area in Fig.~1) reveals a contribution which is hardly detectable at
lower fields because of its larger overlap with the main signal. This secondary signal, which is small (18 \% of the total
$^{59}$Co intensity) and broad (no quadrupolar splitting is resolved), can be singled out at $T$=200~K: At this $T$, the main
signal almost vanishes because it has a shorter spin-spin relaxation time $T_2$. The lower hyperfine shift shows that the
secondary signal comes from less magnetic cobalt ions. All these elements suggest that this signal corresponds to the H1-3 phase
evidenced in XRD data.

\begin{figure}[t!]
\centerline{\includegraphics[width=3.3in]{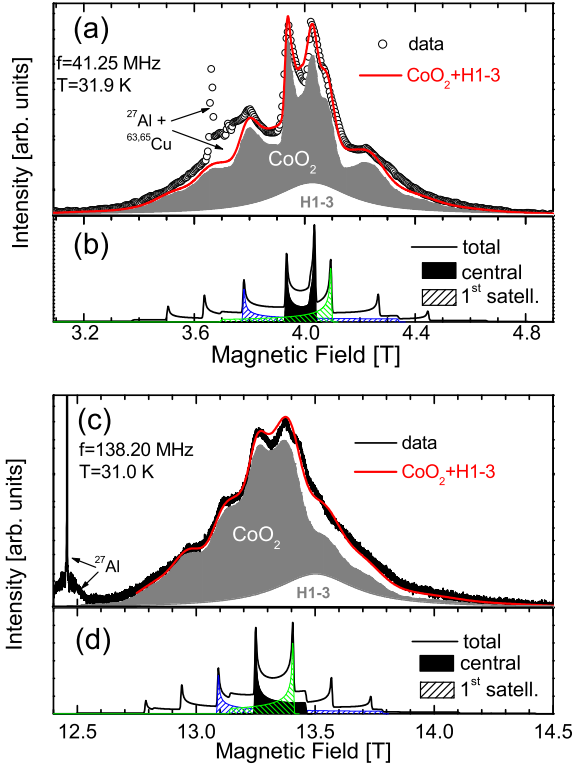}} \vspace{-0.5cm} \caption{(a) and (c): $^{59}$Co NMR powder spectra at two
different frequency values and the simulated spectra with two components (see text). Spurious $^{27}$Al and $^{63,65}$Cu signals
come from the NMR probe. (b) and (d) Corresponding simulations of the CoO$_2$ signal without broadening. For both frequencies,
the lower-field peak corresponds to the central transition of grains oriented with $H\|ab$ (see text). For $H\sim13$~T, the
higher-field peak is mainly the $H\|ab$ part of the first quadrupolar satellite. For $H\sim 4$~T, this peak is primarily caused
by the second order quadrupolar shift on the central line.}
\end{figure}
%
The main signal, attributed to the CoO$_2$ phase, has an asymmetrical shape which is explained by a strong magnetic shift
anisotropy. Since the local symmetry at Co sites is the same as in A$_x$CoO$_2$, we assume axial magnetic symmetry around the $c$
axis, which is also the direction of the principal axis of the electric field gradient, as in all other Na$_x$CoO$_2$
compounds~\cite{Ning04,Mukha05,magneticorder}. Under this assumption, the spectra are perfectly reproduced with a single set of
parameters for the magnetic field values of $\sim$4~T, $\sim$6~T (not shown) and $\sim$13.5~T: The quadrupolar coupling
$\nu_Q$=3.30$\pm$0.05~MHz, the asymmetry parameter $\eta$=0.1$\pm0.1$, and the total hyperfine magnetic shifts
$K_c$=2.35$\pm$0.15~\% and $K_{ab}$=3.95$\pm$0.10~\% (with respect to the resonance frequency of $^{59}$Co nuclei with
$\gamma=10.03$~MHz/T). Each of these parameters, including the linewidth, is constant within error bars over the entire $T$ range
explored (150-1.6~K).

$K_{ab}$ and $K_{c}$ are related to the orbital ($\chi^{\rm orb}$) and the spin ($\chi^{\rm spin}$) susceptibility through
hyperfine couplings $A^{\rm orb}$ and  $A^{\rm spin}$. With the field aligned along $\alpha$, the direction of a principal axis
of both $A$ and $\chi$ tensors,
\begin{equation}
K_{\alpha\alpha} = K^{\rm orb}_{\alpha\alpha} + K^{\rm spin}_{\alpha\alpha}= \frac{A^{\rm orb}_{\alpha\alpha}}{g\mu_B}\chi^{\rm
orb}_{\alpha\alpha} + \frac{A^{\rm spin}_{\alpha\alpha}}{g\mu_B}\chi^{\rm spin}_{\alpha\alpha}.
\end{equation}
$\chi_{\rm bulk}=\chi^{\rm spin}+\chi^{\rm orb}+\chi^{\rm dia}+\chi^{\rm imp}$ (the last two terms stand for the diamagnetic and
impurity contributions, respectively) was measured with a superconducting quantum interference device (SQUID) at a field of
1.45~T (Fig.~2). Below $\sim$70~K, $\chi_{\rm bulk}$ shows a strong increase, the amplitude of which, unlike $^{59}K$ data, is
sample dependent. The amplitude of this rise is also incompatible with $^{59}K$ data: With $^{59}A^{\rm
spin}_{ab}=33$~kOe/$\mu_B$~\cite{CoOH}, the error bars in $^{59}K_{ab}$ translate into a maximum possible variation $\Delta \chi
\simeq 3 \times10^{-4}$~emu/mol from 1.6 to 150~K. The stronger $T$ dependence of $\chi_{\rm bulk}$ at low $T$ is thus attributed
to magnetic impurities or defects. Subtracting a Curie contribution from the raw data restores the weak $T$ dependence imposed by
$K$ data. There remains, however, some uncertainty as to the exact $T$ dependence of the intrinsic $\chi$: As seen in Fig.~2,
subtraction of $\frac{0.0025}{T}$ and $\frac{0.0040}{T}$ terms (representing 0.7 and 1.1 \% of $S$=1/2 moments, respectively)
from $\chi_{\rm bulk}$ results in clear differences below $~$10~K. Furthermore, the 18\% of H1-3 phase, while not expected to
qualitatively influence the discussion, may still slightly affect $\chi_{\rm bulk}$. Despite these uncertainties the data reveal
that the spin susceptibility of CoO$_2$ possesses at most a modest $T$ dependence (if any). A typical $\chi_{\rm spin} =
\frac{c}{T+\theta}$ dependence is ruled out, unless unrealistic values of both $\chi_{\rm orb}$ and $\theta$ are used.

An average $\chi^{\rm spin}\simeq(5\pm1)\times 10^{-4}$~emu/mol is estimated after subtracting $\chi_{\rm dia}+\chi_{\rm
orb}\simeq 2\times 10^{-4}$~emu/mol~\cite{Chou04,Mukha05} from $\chi_{\rm CoO_2}$. $\chi_{\rm spin}$ is thus enhanced by a factor
$\sim$3.6 over the value for non-correlated electrons $\chi_0=2\mu_B^2N(E_F)$, using the LDA-calculated
$N(E_F)=$~4.3~eV$^{-1}$~\cite{theoryU}. In dynamical mean-field theory (DMFT) calculations by Merino {\it et al.} for the
triangular Hubbard model, an enhancement factor $\frac{m^*}{m}\simeq3.6$ corresponds to $U/|t|=9-10$ and $x\simeq0-0.05$ electron
density~\cite{theory}. This suggests the proximity of CoO$_2$ to the Mott insulator transition.

Compared to Na$_{0.3}$CoO$_2$, where $\chi$ is also weakly $T$ dependent without Curie-Weiss behavior, $\chi_{\rm spin}$ in
CoO$_2$ is larger by about $2\times10^{-4}$ emu/mol~\cite{Chou04}. CoO$_2$ might thus be regarded as a Pauli paramagnet with
$\chi^{\rm spin}$ enhanced by electron correlation with respect to $x=0.3$. One should, however, keep in mind that $\chi_{\rm
spin}$ is not the most sensitive quantity for discriminating between various physical descriptions. Antiferromagnetic (AF)
correlations in a triangular lattice lead to a weak $T$ dependence of $\chi^{\rm spin}$, even when moments are localized, and
$\chi^{\rm spin}$ is almost identical in a large $T$ range on both sides of the Mott transition~\cite{Kanoda}. Actually,
spin-lattice relaxation rate ($1/T_1$) data shall establish the correlated-metal nature of CoO$_2$.

$^{59}$Co $T_1$ measurements were performed at the $H\|ab$ position of the central line, and at two magnetic field values: 5.2
and 13.3~T. These experiments probe different ratios of the main to secondary signal amplitudes. Almost identical results in all
the measurements ascertain that the data are intrinsic to the CoO$_2$ phase, with no significant contamination from the secondary
signal. $T_1$ values were extracted from fits of the recovery curves to the appropriate formula for magnetic relaxation. There is
some distribution of $T_1$ values due to mixed contributions from different orientations, as is unavoidable in a random powder.
This can be phenomenologically accounted for by a stretched exponent $\beta$ in the fit of the recovery curves~\cite{Johnston}.
The fit gives $\beta\simeq0.85\pm0.05$, not far from the single-component value $\beta=1$. This is not surprising since the
well-defined $H\|ab$ singularity of the central line largely dominates the total NMR intensity at this position on the spectrum.
The obtained $T_1$ values, identical within errors to those obtained with $\beta=1$, are marginally affected by the distribution.
More importantly, this latter is also strictly $T$ independent, hence it does not affect the obtained $T$ dependence of $T_1$ at
all.

Let us recall that $T_1^{-1}$ is a measurement of low-energy spin fluctuations summed over all wave vectors, with a weighting
factor $A(q)=\sum A^{\rm spin}(r)e^{i\vec{q}.\vec{r}}$:
\begin{equation}
(T_1T)^{-1}=\frac{\gamma_n^2 k_B}{\mu_B^2\hbar}\sum_{q,\alpha\perp H} |A_{\alpha\alpha}(q)|^2
\frac{\chi^{\prime\prime}_{\alpha\alpha}(q,\omega_n)}{\omega_n}
\end{equation}
Above $T^*\simeq7$~K, a fit of $(T_1T)^{-1}$ data to $(\frac{a}{T+\theta})^\alpha$ leads to $\theta=94\pm31$~K and
$\alpha=0.92\pm0.20$. The data thus favor $\alpha=1$, predicted {\it e.g.} for 2D nearly AF metals~\cite{Moriya06}. $\alpha=3/2$
for 2D nearly ferromagnetic metals~\cite{Moriya06} is ruled out while $\alpha=3/4$ calculated for itinerant electrons in a
frustrated 3D lattice~\cite{Lacroix96} cannot be excluded. The data also indicate that any additional $T$ independent
contribution to $(T_1T)^{-1}$, such as orbital relaxation, has to be smaller than 3~s$^{-1}$~K$^{-1}$.
\begin{figure}[t!]
\centerline{\includegraphics[width=3.0in]{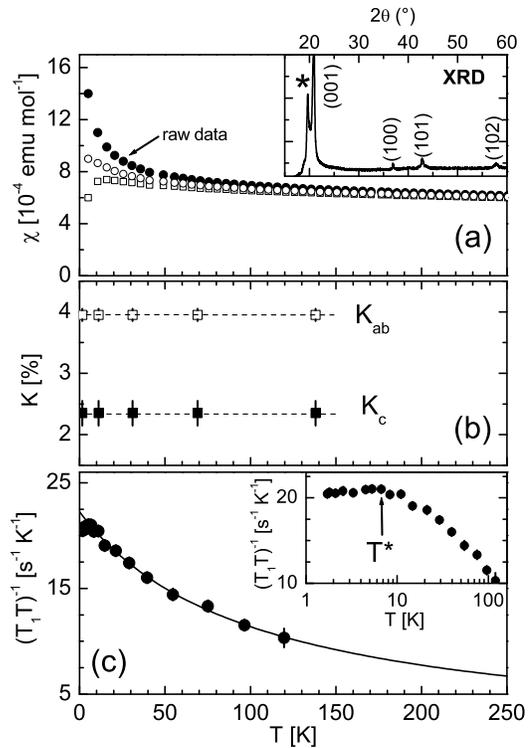}} \vspace{-0.5cm} \caption{(a) SQUID raw data (dots). Open circles and squares
are obtained by subtracting impurity terms $\frac{0.0025}{T}$ and $\frac{0.004}{T}$, respectively, from the raw data (see text).
Inset: XRD data for the O1 CoO$_2$ phase. The peak marked by a star is attributed to the Li-rich H1-3 phase~\cite{H13}. (b)
$^{59}$Co hyperfine shifts $K$ for $H$$\|$$ab$-planes and $H$$\|$$c$-axis with dotted lines to guide the eye. (c) $^{59}$Co
$(T_1T)^{-1}$ data and fit to the $\frac{2200}{T+100}$ dependence. Inset: The same data in log horizontal scale.}
\end{figure}
%
The absence of any marked $T$ dependence of the $q=0$ part of $\chi^\prime_{\rm spin}$, while spin fluctuations are evident from
the Curie-Weiss behavior of $(T_1T)^{-1}$, indicate that spin fluctuations in CoO$_2$ are not ferromagnetic: Correlations at
$q=0$ should show up in an increase of $\chi_{\rm spin}(q=0)$ with decreasing $T$. As a matter of fact, an identical
(Curie-Weiss) $T$ dependence of both $K_{\rm spin}$ and $(T_1T)^{-1}$ is observed in ferromagnetically correlated metals such as
Sr$_{1-x}$Ca$_x$RuO$_3$~\cite{Yoshimura99} or Na$_{0.72}$CoO$_2$~\cite{Ning05}.

Magnetic correlations in CoO$_2$ are (qualitatively) reminiscent of Na$_{0.3}$CoO$_2$$\cdot$$y$H$_2$O~\cite{Ning04,magneticorder}
and of the 2D metal with an anisotropic triangular lattice $\kappa$-Cu[N(CN)$_2$]Br~\cite{Kanoda}. CoO$_2$, however, shows
neither a magnetic nor a superconducting ground state: the clear saturation of $(T_1T)^{-1}$ below $T^*\simeq 7$~K (Inset to
Fig.~2c) suggests that the Korringa law $K^2T_1T$=constant holds, thereby pointing to a Fermi-liquid regime at low $T$. In this
regime, a classical criterion for ferromagnetic (resp. AF) correlations is provided by the value $R<1$ (resp. $R>1$) of the
modified Korringa ratio ${\cal R}=(\frac{\gamma_e}{\gamma_n})^2 \frac{\hbar}{4\pi k_B}(K^2_{\rm
spin}T_1T)^{-1}$~\cite{Korringa1}. In principle, this analysis should consider separately the different hyperfine contributions,
with possible correction factors to ${\cal R}$, such as for example $F_{\rm cp} =\frac{1}{3}f^2+\frac{1}{2}(1-f^2)$ for the core
polarization (with $f$=1 for empty $e_g$ orbitals as in our case)~\cite{Korringa1}. The mechanisms of hyperfine interaction for
$^{59}$Co in the cobaltates are, however, presently not known, making it impossible to disentangle the various contributions.

Separate analysis of $H\|ab$ and $H\|c$ contributions to ${\cal R}$ was not possible in the random powder. However, since
$K_{ab}$ is always larger than $K_c$ in Na$_x$CoO$_2$, the ratio $(\frac{\gamma_e}{\gamma_n})^2 \frac{\hbar}{4\pi
k_B}((K_{ab}^{\rm spin})^2T_1T)^{-1}$ is a lower bound for ${\cal R}$. According to the $x$-dependence of $K^{\rm orb}_{ab}$
revealed by Mukhameshin \etal~\cite{Mukha05}, $K^{\rm orb}_{ab}$ should be larger than 3.44\% in CoO$_2$ (minimum $K^{\rm
orb}_{ab}$ value within error bars for the most magnetic Co$^{\rm +3.7}$ site, corrected for the different shift
reference~\cite{Mukha05}). This implies $K^{\rm spin}_{ab}\leq 0.61$\%, and thus ${\cal R}\geq 2.7$. Alternatively, estimating
$K_{ab}^{\rm spin}\simeq0.3$~\% from Eqn.~1 with $^{59}$A$_{ab}$=33~kOe/$\mu_B$~\cite{CoOH} and $\chi^{\rm spin}\simeq5\times
10^{-4}$~emu/mol, would lead to ${\cal R}$=11. This analysis thus confirms the "antiferromagnetic" ({\it i.e.} $q\neq0$)
character of spin correlations in CoO$_2$.

Finite-$q$ spin fluctuations and low $T$ Korringa behavior are again reminiscent of Na$_{0.3}$CoO$_2$, the lowest-$x$ material
whose magnetic properties have been reported so far~\cite{Ning04,Ning05}. The reduced energy scale $T^* = 7$~K in CoO$_2$,
however, contrasts with $T^*\simeq 80\pm10$~K in Na$_{0.3}$CoO$_2$~\cite{Ning04}. Even if $T^*$ might not be proof of a canonical
Fermi-liquid state~\cite{Wu06}, we interpret the reduced screening temperature $T^*$ in CoO$_2$ as an increase of electronic
correlations as $x\rightarrow 0$. Interestingly, Kondo screening of AF spin fluctuations is consistent with DMFT calculations for
the triangular lattice in the vicinity of the Mott-insulator transition~\cite{theory}.

At this point, it must be stressed that we have obtained consistent results from several CoO$_2$ samples, prepared with the
cleanest possible (electrochemical) method. We also recall that X-ray data rule out any residual amount of Li$^+$ larger than a
few \%, since these would not be accommodated by the specific $x=0$ structure (which phase separates with the $x=0.12$
phase)~\cite{LiNMR}. So, the magnetic properties revealed here for the first time (low $T$ crossover towards a Fermi liquid
ground state with screened spin fluctuations) are intrinsic to CoO$_2$. This behavior, so far unobserved in other members of the
Na$_x$CoO$_2$ family, thus constitutes an important piece of the cobaltate puzzle.

We do not claim, however, that the actual CoO$_2$ material represents ideally undoped Co$^{4+}$ planes. The possibility that
oxygen vacancies make the system electronically equivalent to $x\neq0$ must be considered. The questions of O vacancies and of
the cobalt oxidization state are controversial in Na$_x$CoO$_2$~\cite{Tarascon,Morcrette,Karppinen,Viciu,Mukha05}, so they can
hardly be addressed in CoO$_2$ which is unstable in air and a more complex electrode material. Nevertheless, we recently obtained
similar $T_1$ data in another compound with identical CoO$_2$ layers and a cobalt nominal valence also equal to +4~\cite{CoOH}.
The consistency between these two studies together with the reproducibility of the results in several CoO$_2$ samples strongly
suggest that Co ions are as oxidized as they can be in these materials. Furthermore, from the magnetic point of view, each O
vacancy suppresses pathways for the Co-Co superexchange. In a correlated system with $T$ dependent $\chi(q)$, such a strong
perturbation should cause a $T$-dependent distribution of the local spin polarization~\cite{Julien00}. Therefore, the striking
absence of NMR line broadening on cooling is inconsistent with a large amount of O vacancies in our CoO$_2$ samples. Hence, the
Co oxidization state should be close to +4 within a few \%. Note, however, that this is an effective valence: Most of the
difference between $x=0.3$ and $x=0$ compounds may actually occur in the oxygen band since there is evidence that the oxygen
character of the electronic density increases as $x\rightarrow 0$~\cite{Tarascon,Ning05,xas,chris}.

We thank M.~Morcrette, V.~Simonet and particularly J.-M. Tarascon for their input to the initial part of this project, as well as
M.~Horvati\'c, H.~Mayaffre and R.H. McKenzie for comments and discussions.


\begin{references}
\bibitem{LiNMR} The amount of Li$_{0.12}$CoO$_2$ phase could not be estimated by $^7$Li NMR because
LiPF$_6$ electrolyte residues were not separated from the sample.
\end{references}
\end{document}